\documentclass[a4paper,aps,prd,10pt,preprintnumbers,showpacs,twocolumn,superscriptaddress,nofootinbib,amsmath,amssymb]{revtex4-1}
\usepackage[dvips]{graphics}
\usepackage{cmap}
\usepackage[utf8]{inputenc}
\usepackage[T1]{fontenc}

\def\imo{i}

\def\K{{\cal K}}
\def\Order#1{{\cal O}\left(#1\right)}

\begin{document}
\title{Quasinormal Modes of Dilaton Black Holes: Analytic approximations}
\author{Zainab Malik}\email{zainabmalik8115@gmail.com}
\affiliation{Institute of Applied Sciences and Intelligent Systems, H-15, Pakistan}
\begin{abstract}
We have derived precise analytic expressions for the quasinormal modes of test scalar, and Dirac fields in the background of the dilaton black hole. To achieve this, we employ the higher-order WKB expansion in terms of $1/\ell$. A comparison between the analytic formulas and time-domain integration reveals that the analytic approach generally yields more accurate results than the numerical results previously published using the lower-order WKB approach.
We demonstrate that in the eikonal regime, test fields adhere to the correspondence between null geodesics and eikonal quasinormal modes.
\end{abstract}
\maketitle
\section{Introduction}

Quasinormal modes have been observed using gravitational interferometers operated by the LIGO/VIRGO collaboration \cite{LIGOScientific:2016aoc,LIGOScientific:2017vwq,LIGOScientific:2020zkf}. The planned experiments hold the promise of a much broader frequency band for observations \cite{Babak:2017tow}. When combined with ongoing and future observations of black holes in the electromagnetic spectrum \cite{EventHorizonTelescope:2019dse,Goddi:2016qax}, these endeavors could potentially impose stronger constraints on black hole geometry \cite{Tsukamoto:2014tja,Shaikh:2021yux}.
Simultaneously, the uncertainty in determining the mass and angular momentum of the final black hole allows for associating the observed signal with a broad class of non-Kerr spacetimes, thereby enabling exploration of a range of alternative theories of gravity.
In this context, our focus is on a particular solution obtained as a low-energy limit of string theory within the framework of the Einstein-Maxwell-dilaton theory \cite{Gibbons:1987ps,Garfinkle:1990qj}.

Typically, quasinormal frequencies of black holes are determined numerically due to the inherent complexity of the master wave-like equation. There are only a few exceptions, such as some lower-dimensional black holes like the $(2+1)$-dimensional BTZ spacetime \cite{Banados:1992wn} and some of its generalizations. Even for the simplest $(3+1)$ Schwarzschild black hole, an exact analytic solution remains elusive. Additionally, certain $(2+1)$-dimensional generalizations of the BTZ spacetimes \cite{Konoplya:2020ibi} necessitate numerical methods \cite{Skvortsova:2023zmj} because the metric functions are no longer simple. Another exception comes in the form of approximate analytic expressions for quasinormal modes in specific parameter regimes, such as the near-extreme asymptotically de Sitter spacetime \cite{Cardoso:2003sw,Molina:2003ff} or the high multipole number eikonal regime. Although the eikonal formula lacks practical value due to its inaccuracy at smaller multipole numbers $\ell$, it surprisingly produces precise expressions when extended beyond the eikonal regime \cite{Konoplya:2023moy}. Therefore, in this work, we adopt the latter approach to derive the analytic form of quasinormal modes for a dilaton black hole.

Our work places particular emphasis on the eikonal regime, characterized by high multipole numbers or, equivalently, high real oscillation frequencies. This regime is of significant interest for several reasons. Firstly, a correspondence between eikonal quasinormal frequencies and null geodesics, formulated in \cite{Cardoso:2008bp}, establishes connections between the real and imaginary parts of the eikonal quasinormal frequencies and the rotational frequency and Lyapunov exponent of the unstable circular null geodesic, respectively. Subsequent studies have revealed exceptions to this correspondence \cite{Khanna:2016yow,Konoplya:2019hml,Bolokhov:2023dxq}, and the true limits of its applicability have been discussed in \cite{Konoplya:2017wot,Konoplya:2022gjp,Bolokhov:2023dxq}. It was demonstrated that the correspondence holds only for the part of the spectrum, if any, that can be reproduced by the WKB formula of \cite{Mashhoon:1982im,Schutz:1985km}.

Moreover, the eikonal regime may unveil catastrophic and unusual instabilities \cite{Takahashi:2011du,Takahashi:2011qda,Takahashi:2010gz,Dotti:2004sh,Dotti:2005sq,Gleiser:2005ra,Cuyubamba:2016cug}, particularly in theories with higher curvature corrections, such as Einstein-Gauss-Bonnet \cite{Dotti:2004sh,Dotti:2005sq,Gleiser:2005ra} or Einstein-Lovelock \cite{Konoplya:2017lhs}. The crux of this effect lies in the stability of lower multipole numbers contrasted with the instability introduced by higher ones, which manifests after a prolonged period of damped oscillations.

With the aforementioned motivations in mind, eikonal formulas for quasinormal modes have been derived for perturbations in various theories of gravity and diverse black hole models. Specifically, in D-dimensional Einstein theory, the eikonal formula was deduced \cite{Konoplya:2003ii}. For Einstein-Weyl gravity, the corresponding formula was derived in \cite{Kokkotas:2017zwt}. In the case of Einstein-dilaton-Gauss-Bonnet theory, the eikonal expression was established in \cite{Konoplya:2019hml}, and for the quantum-corrected Kazakov-Solodukhin black hole, it was obtained in \cite{Konoplya:2019xmn}. Additionally, eikonal expressions for the four-dimensional Einstein-Gauss-Bonnet theory were deduced in \cite{Konoplya:2020bxa}, and for the holonomy quantum-corrected theory, it was formulated in \cite{Bolokhov:2023bwm}. Further instances include the eikonal expression for Schwarzschild-de Sitter spacetime \cite{Zhidenko:2003wq} and for the dynamical Chern-Simons theory \cite{Chen:2022nlw}. Eikonal formulas were also established for scalar-tensor theories \cite{Silva:2019scu,Glampedakis:2019dqh}, the Visser-Simpson black hole model \cite{Churilova:2019cyt}, non-linear electrodynamics \cite{Toshmatov:2019gxg}, and black holes with quadrupole momentum \cite{Allahyari:2018cmg}. The general approach to finding eikonal formulas for spherically symmetric black holes was explored in \cite{Glampedakis:2019dqh,Churilova:2019jqx}, with an extension beyond the eikonal regime detailed in \cite{Konoplya:2023moy}.

The particular class of black holes we are interested in this paper is dilatonic black holes, which are solutions to the Einstein-Maxwell-Dilaton (EMD) equations. The dilaton field is a scalar field originating from string theory, representing the geometry of spacetime. Thereby, a dilatonic black hole is a model for the quantum corrected black hole. In this paper, we aim to derive the eikonal formula for scalar and Dirac perturbations around the dilatonic black hole. Moreover, we extend the analytic formula to a few orders beyond the eikonal approximation, resulting in a relatively accurate analytic expression for the quasinormal frequencies of dilatonic black holes. Our analysis demonstrates that the obtained analytic expression surpasses the accuracy of certain previously published numerical data. Additionally, we conduct tests to validate the correspondence between eikonal quasinormal modes and null geodesics.

The paper is structured as follows: In Section II, we provide the fundamental expressions for the metric and the wave-like equations. Section III elucidates the WKB approach, while in Section IV, we derive the eikonal and beyond-eikonal analytic formulas for quasinormal modes, comparing them with data obtained through other methods. Finally, in Section V, we summarize the findings of our study.

\section{Basic equations}\label{sec:wavelike}

We will deal with the Einstein-Maxwell-dilaton theory, which describe coupling gravitational, electromagnetic and scalar
fields with the action:
\begin{equation}
S = \int d^{4} x \sqrt{-g} (\mathcal{R} - 2 (\nabla \Phi)^{2} + e^{- 2 \Phi} F^{2}),
\end{equation}
where $\mathcal{R}$ is the curvature, $F$ is the electromagnetic tensor and $\Phi$ is the scalar (dilaton) field.

The metric of the dilaton black hole is given by the following line element \cite{Gibbons:1987ps,Garfinkle:1990qj,Garcia:1995qz},
\begin{equation}\label{metric}
  ds^2=-f(r)dt^2+\frac{dr^2}{f(r)}+R^2(r)(d\theta^2+\sin^2\theta d\phi^2),
\end{equation}
where
$$
\begin{array}{rcl}
f(r)&=&\displaystyle \frac{2 b-2 M+r}{2 b+r},\\
R^2(r)&=&\displaystyle r (2 b+r),\\
\end{array}
$$
where $b$ is the dilaton parameter, and $M$ is the ADM mass. We shall further measure all dimensional quantities in units of the mass, i.~e., we choose $M=1$.

The general relativistic equations for the scalar ($\Phi$),
and Dirac ($\Upsilon$) fields can be written in the following form:
\begin{subequations}\label{coveqs}
\begin{eqnarray}\label{KGg}
\frac{1}{\sqrt{-g}}\partial_\mu \left(\sqrt{-g}g^{\mu \nu}\partial_\nu\Phi\right)&=&0,
\\\label{covdirac}
\gamma^{\alpha} \left( \frac{\partial}{\partial x^{\alpha}} - \Gamma_{\alpha} \right) \Upsilon&=&0,
\end{eqnarray}
\end{subequations}
where, $\gamma^{\alpha}$ are noncommutative gamma matrices and $\Gamma_{\alpha}$ are spin connections in the tetrad formalism.
After separation of the variables in the background (\ref{metric}) the above equations (\ref{coveqs}) take the Schrödinger wavelike form \cite{Kokkotas:1999bd,Berti:2009kk,Konoplya:2011qq}:
\begin{equation}\label{wave-equation}
\dfrac{d^2 \Psi}{dr_*^2}+(\omega^2-V(r))\Psi=0,
\end{equation}
where the ``tortoise coordinate'' $r_*$ is defined as follows:
\begin{equation}\label{tortoise}
dr_*\equiv\frac{dr}{f(r)}.
\end{equation}

The effective potential for the scalar field has the form
\begin{equation}\label{potentialScalar}
V(r)=f(r)\frac{\ell(\ell+1)}{R(r)^2}+\frac{1}{R(r)}\cdot\frac{d^2 R(r)}{dr_*^2},
\end{equation}
where $\ell=0, 1, 2, \ldots$ are the multipole numbers.
For the Dirac field there are two isospectral potentials,
\begin{equation}
V_{\pm}(r) = W^2\pm\frac{dW}{dr_*}, \quad W\equiv \left(\ell+\frac{1}{2}\right)\frac{\sqrt{f(r)}}{R(r)}.
\end{equation}
The isospectral wave functions can be transformed one into another by the Darboux transformation,
\begin{equation}\label{psi}
\Psi_{+}\propto \left(W+\dfrac{d}{dr_*}\right) \Psi_{-},
\end{equation}
so that it is sufficient to calculate quasinormal modes for only one of the effective potentials. We will do that for $V_{+}(r)$ because from the earlier publications on Dirac quasinormal modes \cite{Zinhailo:2019rwd} it is known that the WKB method works better in this case. The effective potentials are shown in figs. 1-3.

Quasinormal modes of various dilaton black hole were considered in a great number of works  \cite{Ferrari:2000ep,Konoplya:2001ji,Carson:2020ter,Malybayev:2021lfq,Konoplya:2022zav,Pani:2009wy,Konoplya:2019hml,Lopez-Ortega:2009jpx,Kokkotas:2017ymc,Fernando:2003wc,Chen:2005rm,Lopez-Ortega:2005obq}, though there the results were presented in the numerical form only.

\begin{figure}
\resizebox{\linewidth}{!}{\includegraphics{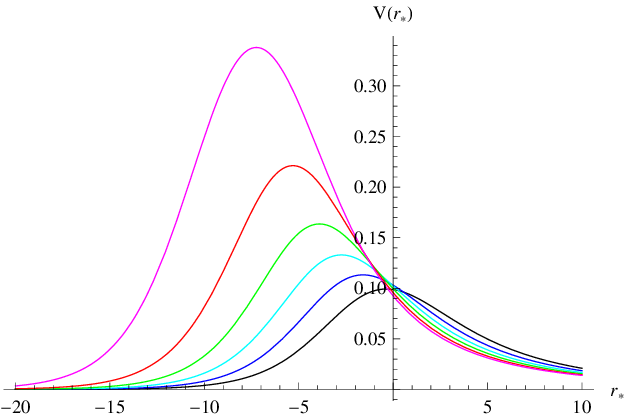}}
\caption{Plot of the effective potential for the $\ell=1$ test scalar field: $b=0$ (black), $b=0.2$ (blue), $b=0.4$ (cyan), $b=0.6$ (green), $b=0.8$ (red), $b=0.95$ (magenta).}\label{fig:scalarpot}
\end{figure}

\begin{figure}
\resizebox{\linewidth}{!}{\includegraphics{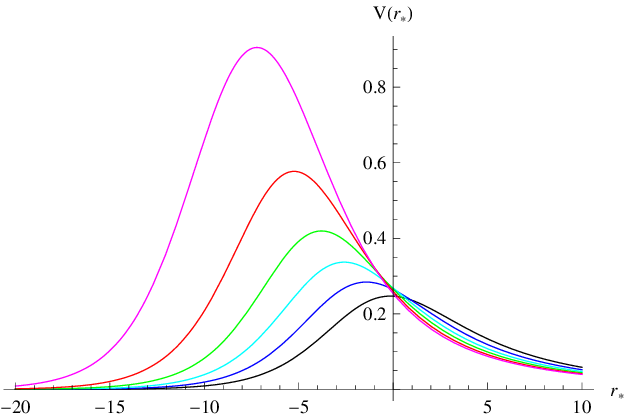}}
\caption{Plot of the effective potential for the $\ell=2$ test scalar field: $b=0$ (black), $b=0.2$ (blue), $b=0.4$ (cyan), $b=0.6$ (green), $b=0.8$ (red), $b=0.95$ (magenta).}\label{fig:l2pot}
\end{figure}

\begin{figure*}
\resizebox{\linewidth}{!}{\includegraphics{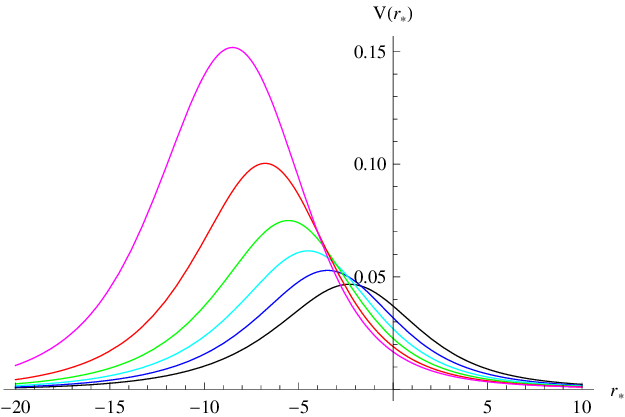}\includegraphics{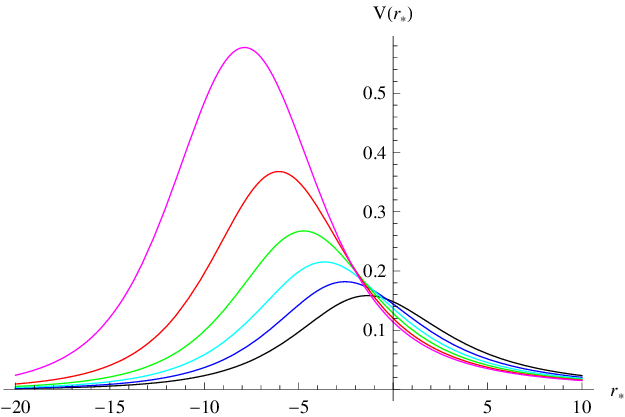}}
\caption{Plot of the effective potential for the Dirac field $\ell=1/2$ (left panel) and $\ell=3/2$ (right panel): $b=0$ (black), $b=0.2$ (blue), $b=0.4$ (cyan), $b=0.6$ (green), $b=0.8$ (red), $b=0.95$ (magenta).}\label{fig:Diracpot}
\end{figure*}

\section{WKB formula}\label{sec:WKB}

When the effective potential $V(r)$ in the wavelike equation (\ref{wave-equation}), has a form of a barrier with a single peak, the WKB formula is appropriate for obtaining the dominant quasinormal modes, satisfying the boundary conditions,
\begin{equation}\label{boundaryconditions}
\Psi(r_*\to\pm\infty)\propto e^{\pm\imo \omega r_*},
\end{equation}
which are purely ingoing wave at the horizon ($r_*\to-\infty$) and purely outgoing wave at spatial infinity ($r_*\to\infty$).
\par

The WKB method is based on matching of the two asymptotic solutions, at the event horizon and infinity, which satisfy the quasinormal boundary conditions (\ref{boundaryconditions}), with the Taylor expansion around the peak of the potential barrier. The first-order WKB formula represents the eikonal approximation and is exact in the limit $\ell \to \infty$. Then, the general WKB expression for the frequencies can be written in the form of expansion around the eikonal limit as follows \cite{Konoplya:2019hlu}:
\begin{eqnarray}\label{WKBformula-spherical}
\omega^2&=&V_0+A_2(\K^2)+A_4(\K^2)+A_6(\K^2)+\ldots\\\nonumber&-&\imo \K\sqrt{-2V_2}\left(1+A_3(\K^2)+A_5(\K^2)+A_7(\K^2)\ldots\right),
\end{eqnarray}
and the matching conditions for the quasinormal modes imply that
\begin{equation}
\K=n+\frac{1}{2}, \quad n=0,1,2,\ldots,
\end{equation}
where $n$ is the overtone number, $V_0$ is the maximal value of the effective potential, $V_2$ is the value of the second derivative of the potential in this point with respect to the tortoise coordinate, and $A_i$ for $i=2, 3, 4, \ldots$ is $i-th$ WKB order correction term beyond the eikonal approximation which depends on $\K$ and derivatives of the potential in its maximum up to the order $2i$. The explicit form of $A_i$ can be found in \cite{Iyer:1986np} for the second and third WKB order, in \cite{Konoplya:2003ii} for the 4-6th orders and in \cite{Matyjasek:2017psv} for the 7-13th orders.
While, strictly speaking, the WKB series converges only asymptotically, usually it gives unprecedented accuracy of the results for the low-lying modes, especially when using the Padé approximants as shown in \cite{Matyjasek:2017psv}. In the present paper when comparing our analytical formula with the 6th order WKB method, we imply the latter modification of it with the Padé approximants.
Therefore, the above WKB approach was used for finding of quasinormal modes and grey-body factors at various orders in a great number of works
(see, for example, \cite{Konoplya:2005sy,Ishihara:2008re,Konoplya:2018ala,Kodama:2009bf,Yang:2023gas,Fernando:2012yw,Karmakar:2023cwg,Barrau:2019swg,Xia:2023zlf} and references therein).

\begin{table}
\begin{tabular}{ c c c c }
\hline
b & WKB6 & analytic & error \\
\hline
$0$ & $0.2929-0.0978 i$ & $0.2928-0.0977 i$ & $0.027\%$ \\
$0.10$ & $0.3034-0.0988 i$ & $0.3033-0.0988 i$ & $0.028\%$ \\
$0.20$ & $0.3152-0.0999 i$ & $0.3151-0.0999 i$ & $0.045\%$ \\
$0.30$ & $0.3287-0.1011 i$ & $0.3284-0.1011 i$ & $0.111\%$ \\
$0.40$ & $0.3444-0.1023 i$ & $0.3434-0.1024 i$ & $0.301\%$ \\
$0.50$ & $0.3631-0.1036 i$ & $0.3603-0.1037 i$ & $0.738\%$ \\
$0.60$ & $0.38590-0.10483 i$ & $0.37940-0.10515 i$ & $1.63\%$ \\
$0.70$ & $0.41506-0.10594 i$ & $0.40086-0.10668 i$ & $3.32\%$ \\
$0.80$ & $0.45509-0.10643 i$ & $0.42491-0.10831 i$ & $6.47\%$ \\
$0.85$ & $0.48235-0.10601 i$ & $0.43796-0.10916 i$ & $9.01\%$ \\
$0.90$ & $0.51863-0.10438 i$ & $0.4517-0.11003 i$ & $12.7\%$ \\
$0.95$ & $0.5737-0.0992 i$ & $0.4663-0.1109 i$ & $18.6\%$ \\
\hline
\end{tabular}
\caption{Quasinormal modes of the $\ell=1$ test scalar field for the dilaton black hole calculated using the 6th order WKB formula and the approximate analytic formula. The deviation is given in per cents.}
\end{table}

\begin{table}
\begin{tabular}{ c c c c }
\hline
b & WKB6 & analytic & error \\
\hline
$0$ & $0.48364-0.09677 i$ & $0.48362-0.09677 i$ & $0.0045\%$ \\
$0.10$ & $0.50080-0.09785 i$ & $0.50077-0.09785 i$ & $0.0052\%$ \\
$0.20$ & $0.52018-0.09901 i$ & $0.52007-0.09901 i$ & $0.0195\%$ \\
$0.30$ & $0.54234-0.10025 i$ & $0.54187-0.10025 i$ & $0.0858\%$ \\
$0.40$ & $0.56813-0.10155 i$ & $0.56652-0.10157 i$ & $0.279\%$ \\
$0.50$ & $0.59878-0.10291 i$ & $0.59437-0.10299 i$ & $0.726\%$ \\
$0.60$ & $0.63633-0.10428 i$ & $0.62579-0.10449 i$ & $1.63\%$ \\
$0.70$ & $0.68437-0.10551 i$ & $0.66112-0.10610 i$ & $3.36\%$ \\
$0.80$ & $0.750420-0.106156 i$ & $0.700717-0.107812 i$ & $6.56\%$ \\
$0.85$ & $0.795440-0.105823 i$ & $0.722227-0.108709 i$ & $9.13\%$ \\
$0.90$ & $0.85544-0.10428 i$ & $0.7449-0.10963 i$ & $12.8\%$ \\
$0.95$ & $0.94669-0.09910 i$ & $0.76889-0.11059 i$ & $18.7\%$ \\
\hline
\end{tabular}
\caption{Quasinormal modes of the $\ell=2$ test scalar field for the dilaton black hole calculated using the 6th order WKB formula and the approximate analytic formula. The deviation is given in per cents.}
\end{table}

\begin{table}
\begin{tabular}{ c c c c }
\hline
b & WKB6 & analytic & error \\
\hline
$0$ & $0.1826-0.0949 i$ & $0.1826-0.0969 i$ & $0.975\%$ \\
$0.10$ & $0.1896-0.0963 i$ & $0.1897-0.0980 i$ & $0.826\%$ \\
$0.20$ & $0.1974-0.0977 i$ & $0.1976-0.0992 i$ & $0.678\%$ \\
$0.30$ & $0.2064-0.0993 i$ & $0.2066-0.1005 i$ & $0.528\%$ \\
$0.40$ & $0.2169-0.1009 i$ & $0.2167-0.1018 i$ & $0.388\%$ \\
$0.50$ & $0.2294-0.1026 i$ & $0.2282-0.1033 i$ & $0.556\%$ \\
$0.60$ & $0.2449-0.1044 i$ & $0.2411-0.1049 i$ & $1.41\%$ \\
$0.70$ & $0.2648-0.1060 i$ & $0.2557-0.1066 i$ & $3.20\%$ \\
$0.80$ & $0.2927-0.1070 i$ & $0.2721-0.1085 i$ & $6.62\%$ \\
$0.85$ & $0.3119-0.1068 i$ & $0.2810-0.1095 i$ & $9.41\%$ \\
$0.90$ & $0.3379-0.1052 i$ & $0.290-0.1105 i$ & $13.5\%$ \\
$0.95$ & $0.3779-0.1000 i$ & $0.3003-0.1115 i$ & $20.1\%$ \\
\hline
\end{tabular}
\caption{Quasinormal modes of the $\ell=1/2$ Dirac field for the dilaton black hole calculated using the 6th order WKB formula and the approximate analytic formula. The deviation is given in per cents.}
\end{table}

\begin{table}
\begin{tabular}{ c c c c }
\hline
b & WKB6 & analytic & error \\
\hline
$0$ & $0.3801-0.0964 i$ & $0.3800-0.0964 i$ & $0.020\%$ \\
$0.10$ & $0.3938-0.0975 i$ & $0.3937-0.0975 i$ & $0.021\%$ \\
$0.20$ & $0.4093-0.0986 i$ & $0.4091-0.0987 i$ & $0.032\%$ \\
$0.30$ & $0.42700-0.09988 i$ & $0.42659-0.09995 i$ & $0.0954\%$ \\
$0.40$ & $0.44765-0.10122 i$ & $0.44633-0.10130 i$ & $0.289\%$ \\
$0.50$ & $0.47223-0.10264 i$ & $0.46865-0.10276 i$ & $0.741\%$ \\
$0.60$ & $0.50237-0.10408 i$ & $0.49384-0.10431 i$ & $1.66\%$ \\
$0.70$ & $0.54102-0.10541 i$ & $0.52218-0.10598 i$ & $3.42\%$ \\
$0.80$ & $0.59429-0.10622 i$ & $0.55396-0.10777 i$ & $6.69\%$ \\
$0.85$ & $0.63070-0.10601 i$ & $0.57123-0.10870 i$ & $9.31\%$ \\
$0.90$ & $0.67934-0.10463 i$ & $0.5895-0.10967 i$ & $13.1\%$ \\
$0.95$ & $0.75357-0.09969 i$ & $0.60870-0.11068 i$ & $19.1\%$ \\
\hline
\end{tabular}
\caption{Quasinormal modes of the $\ell=3/2$ Dirac field for the dilaton black hole calculated using the 6th order WKB formula and the approximate analytic formula. The deviation is given in per cents.}
\end{table}

\section{Eikonal limit and analytic expression beyond eikonal}

\begin{figure}
\resizebox{\linewidth}{!}{\includegraphics{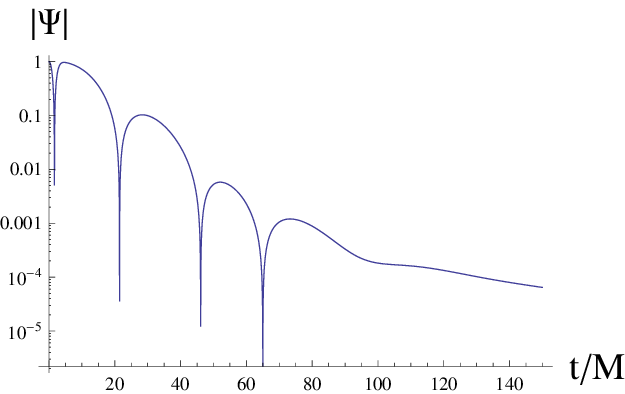}}
\caption{Time-domain profile for the scalar field ($\ell=0$) in the background of the dilatonic black hole ($b=0.5M$).}\label{fig:timedomain0}
\end{figure}

\begin{figure}
\resizebox{\linewidth}{!}{\includegraphics{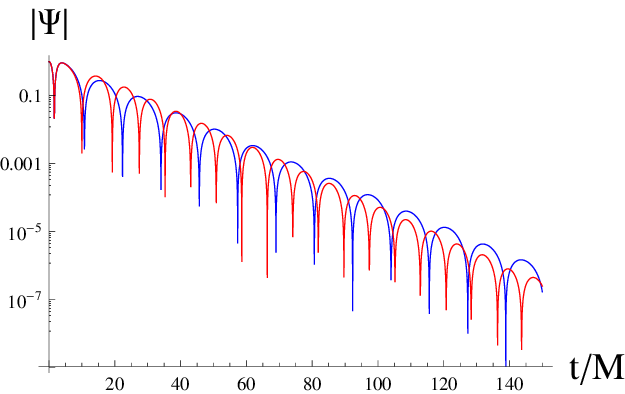}}
\caption{Time-domain profile for the scalar field ($\ell=1$) in the background of the dilatonic black hole $b=0.2M$ (blue) and $b=0.8M$ (red).}\label{fig:timedomain1}
\end{figure}

\begin{table*}
\begin{tabular}{c c c c c c}
\hline
$b$ & Prony fit & analytic & error & WKB3 & error\\
\hline
$0$ & $0.11-0.106 i$ & $0.10870-0.10979 i$ & $2.62\%$ & $0.1046-0.1152 i$ & $6.97\%$\\
$0.2$ & $0.12-0.109 i$ & $0.11764-0.11099 i$ & $1.91\%$ & $0.1140-0.1156 i$ & $5.49\%$\\
$0.5$ & $0.145-0.118 i$ & $0.13541-0.11281 i$ & $5.83\%$ & $0.1346-0.1164 i$ & $5.64\%$\\
$0.9$ & $0.206-0.115 i$ & $0.17052-0.11505 i$ & $15.0\%$ & $0.2160-0.1210 i$ & $4.92\%$\\
\hline
\end{tabular}
\caption{Comparison of the time-domain fit and analytic formula for $\ell=0$ scalar field perturbations ($M=1$).}
\end{table*}

\begin{table*}
\begin{tabular}{c c c c c c}
\hline
$b$ & Prony fit & analytic & error & WKB3 & error\\
\hline
$0.2$ & $0.3152-0.0997 i$ & $0.31507-0.09990 i$ & $0.0719\%$ & $0.3137-0.1001 i$ & $0.483\%$\\
$0.4$ & $0.3443-0.1022 i$ & $0.34335-0.10236 i$ & $0.267\%$ & $0.3432-0.1025 i$ & $0.324\%$\\
$0.6$ & $0.3857-0.1042 i$ & $0.37940-0.10515 i$ & $1.60\%$ & $0.3851-0.1052 i$ & $0.280\%$\\
$0.8$ & $0.4514-0.1023 i$ & $0.42491-0.10831 i$ & $5.87\%$ & $0.4551-0.1069 i$ & $1.28\%$\\
\hline
\end{tabular}
\caption{Comparison of the time-domain fit and analytic formula for $\ell=1$ scalar field perturbations ($M=1$).}
\end{table*}

Perturbations in a spherically symmetric background can be reduced to the wave-like equation with the effective potential which can be approximated in the following way:
\begin{equation}\label{potential-multipole}
V(r_*)=\kappa^2\left(H(r_*)+\Order{\kappa^{-1}}\right).
\end{equation}
Here $\kappa\equiv\ell+\frac{1}{2}$ and $\ell=s,s+1,s+2,\ldots$ is the positive half(integer) multipole number, which has minimal value equal to the spin of the field under consideration $s$. Here, following \cite{Konoplya:2023moy} we use an expansion in terms of $\kappa^{-1}$.

The function $H(r_*)$ has a single peak, so that, the location of the potential's maximum (\ref{potential-multipole}) can be expanded as follows
\begin{equation}\label{rmax}
  r_{\max }=r_0+r_1\kappa^{-1}+r_2\kappa^{-2}+\ldots.
\end{equation}

Substituting (\ref{rmax}) into the following first order WKB formula
\begin{eqnarray}
\omega&=&\sqrt{V_0-\imo \K\sqrt{-2V_2}},
\end{eqnarray}
and then expanding in $\kappa^{-1}$, we find that,
\begin{eqnarray}\label{eikonal-formulas}
\omega=\Omega\kappa-\imo\lambda\K+\Order{\kappa^{-1}}.
\end{eqnarray}
The above relation is a reasonable approximation for $\kappa\gg\K$.

In \cite{Cardoso:2008bp} it was shown that parameters of the unstable circular null geodesics around a static and spherically symmetric  black hole (asymptotically flat or de Sitter one) are dual to the quasinormal modes emitted by the black hole in the $\ell \gg n$ regime: the real and imaginary components of the $\ell \gg n$ quasinormal frequency are proportional to the frequency and instability timescale of the circular null geodesics, given by:
\begin{equation}\label{QNM}
\omega_n=\Omega\ell-\imo(n+1/2)|\lambda|, \quad \ell \gg n.
\end{equation}
Here $\Omega$ is the angular velocity at the unstable null geodesics, and $\lambda$ is the Lyapunov exponent.

The radius of the unstable circular null geodesic $r_{inst}$ occurs at $U'(r)=0$, where $U(r)$ is the effective potential for a circular null geodesic. Therefore, it satisfies the relation
\begin{equation}\label{radius}
2 f(r_{inst}) = r_{inst} f'(r)_{r=r_{inst}},
\end{equation}
while the angular velocity is
\begin{equation}\label{omega}
\Omega=\frac{d \phi} {d t} = \frac{f(r_{inst})^{1/2}}{r_{inst}}.
\end{equation}
In the eikonal regime $r_{inst} = r_{max}$.

While this correspondence holds for a great number of cases, in \cite{Konoplya:2017wot} it was shown that it breaks down when the usual centrifugal term $f(r) \ell (\ell +1)/r^2$ in the effective potential has a different form, as, for example, in the Einstein-Gauss-Bonnet theories allowing, generally, for a scalar field \cite{Konoplya:2017wot,Konoplya:2020bxa,Konoplya:2019hml}. While it was demonstrated that the basically the correspondence works, while the WKB formula of \cite{Schutz:1985km} for quasinormal modes is valid  \cite{Konoplya:2017wot}, later it tuned out that even in those cases not the full eikonal spectrum might be reproduced via the WKB formula and, consequently, not the overtones number $n$ may not correspond to the real number of the frequency \cite{Konoplya:2022gjp,Bolokhov:2023dxq}.

An examination of the aforementioned analytic formulas in the eikonal limit and the above expressions for the angular velocity and Lyapunov exponent reveals that the correspondence between the eikonal quasinormal modes and null geodesics is established. This fulfillment arises because the eikonal quasinormal modes are strictly calculated based on the first-order WKB formula, and in the eikonal regime, the peak of the potential barrier coincides with the photon orbit.

In principle, the expansion (\ref{rmax}) can be extended to an arbitrary order, allowing for the determination of higher-order corrections to the eikonal formula (\ref{eikonal-formulas}). This involves applying the higher-order WKB formula, where the order $n$ WKB formula can be utilized to identify corrections of the order $\kappa^{1-n}$ to the analytic expression (\ref{eikonal-formulas}).

\begin{widetext}
Using the series expansion in terms of the inverse multipole number \cite{Konoplya:2023moy}, for the scalar field we find the expansion for the location of the potential peak,
\begin{equation}\label{rmax-scalar}
r_{\max } = 3 M-\frac{M}{3 \kappa ^2}+b \left(\frac{10}{27 \kappa ^2}-\frac{7}{3}\right)+b^2 \left(-\frac{2}{81 M \kappa ^2}-\frac{4}{27 M}\right)+b^3 \left(-\frac{14}{2187 M^2 \kappa ^2}-\frac{20}{243 M^2}\right)+\mathcal{O}\left(b^4,\frac{1}{\kappa ^4}\right)
\end{equation}
and, using the WKB formula, the expression
\begin{equation}\label{eikonal-scalar}
\begin{array}{rcl}
\omega  &=& \displaystyle-\frac{i K \left(940 K^2+313\right)}{46656 \sqrt{3} M \kappa ^2}+\frac{29-60 K^2}{1296 \sqrt{3} M \kappa }+\frac{\kappa }{3 \sqrt{3} M}-\frac{i K}{3 \sqrt{3} M}\\
&&\displaystyle+b \left(\frac{5 i K \left(220 K^2+61\right)}{139968 \sqrt{3} M^2 \kappa ^2}+\frac{204 K^2+31}{11664 \sqrt{3} M^2 \kappa }+\frac{\kappa }{9 \sqrt{3} M^2}-\frac{i K}{27 \sqrt{3} M^2}\right)\\
&&\displaystyle+b^2 \left(\frac{i K \left(1324 K^2+697\right)}{419904 \sqrt{3} M^3 \kappa ^2}+\frac{588 K^2+47}{104976 \sqrt{3} M^3 \kappa }+\frac{5 \kappa }{81 \sqrt{3} M^3}-\frac{i K}{81 \sqrt{3} M^3}\right)\\
&&\displaystyle+b^3 \left(\frac{i K \left(47468 K^2+41705\right)}{34012224 \sqrt{3} M^4 \kappa ^2}+\frac{4596 K^2-119}{944784 \sqrt{3} M^4 \kappa }+\frac{89 \kappa }{2187 \sqrt{3} M^4}-\frac{7 i K}{2187 \sqrt{3} M^4}\right)+\mathcal{O}\left(b^4,\frac{1}{\kappa ^3}\right)
\end{array}
\end{equation}
for the quasinormal modes (where $\kappa\equiv\ell+1/2$, $K\equiv n+1/2$).

Finally, for the Dirac field, we find
\begin{equation}\label{rmax-Dirac}
\begin{array}{rcl}
r_{\max } &=& \displaystyle\frac{11 M}{16 \sqrt{3} \kappa ^3}-\frac{\sqrt{3} M}{2 \kappa }+3 M\\
&&\displaystyle+b \left(-\frac{95}{144 \sqrt{3} \kappa ^3}+\frac{1}{9 \kappa ^2}+\frac{5}{6 \sqrt{3} \kappa }-\frac{7}{3}\right)\\
&&\displaystyle+b^2 \left(-\frac{25}{648 \sqrt{3} M \kappa ^3}-\frac{1}{27 M \kappa ^2}+\frac{5}{27 \sqrt{3} M \kappa }-\frac{4}{27 M}\right)\\
&&\displaystyle+b^3 \left(-\frac{19}{1944 \sqrt{3} M^2 \kappa ^3}-\frac{11}{729 M^2 \kappa ^2}+\frac{7}{81 \sqrt{3} M^2 \kappa }-\frac{20}{243 M^2}\right)+\mathcal{O}\left(b^4,\frac{1}{\kappa ^4}\right)
\end{array}
\end{equation}
and the formula for the quasinormal modes reads
\begin{equation}\label{eikonal-Dirac}
\begin{array}{rcl}
\omega  &=& \displaystyle\frac{i K \left(119-940 K^2\right)}{46656 \sqrt{3} M \kappa ^2}-\frac{60 K^2+7}{1296 \sqrt{3} M \kappa }+\frac{\kappa }{3 \sqrt{3} M}-\frac{i K}{3 \sqrt{3} M}\\
&&\displaystyle+b \left(\frac{i K \left(1100 K^2-127\right)}{139968 \sqrt{3} M^2 \kappa ^2}+\frac{204 K^2-5}{11664 \sqrt{3} M^2 \kappa }+\frac{\kappa }{9 \sqrt{3} M^2}-\frac{i K}{27 \sqrt{3} M^2}\right)\\
&&\displaystyle+b^2 \left(\frac{i K \left(1324 K^2-887\right)}{419904 \sqrt{3} M^3 \kappa ^2}+\frac{588 K^2+83}{104976 \sqrt{3} M^3 \kappa }+\frac{5 \kappa }{81 \sqrt{3} M^3}-\frac{i K}{81 \sqrt{3} M^3}\right)\\
&&\displaystyle+b^3 \left(\frac{i K \left(47468 K^2-62407\right)}{34012224 \sqrt{3} M^4 \kappa ^2}+\frac{4596 K^2+925}{944784 \sqrt{3} M^4 \kappa }+\frac{89 \kappa }{2187 \sqrt{3} M^4}-\frac{7 i K}{2187 \sqrt{3} M^4}\right)+\mathcal{O}\left(b^4,\frac{1}{\kappa ^3}\right)
\end{array}
\end{equation}

\end{widetext}

The accuracy of the above analytic formulas can be checked here in two ways. First, by comparison with the 6th order WKB formula with Padé approximants and, what is more independent approach, with the time-domain integration. Here for the integration in time domain we used the Gundlach-Price-Pullin discretization scheme \cite{Gundlach:1993tp}
\begin{eqnarray}\label{Discretization}
\Psi\left(N\right)&=&\Psi\left(W\right)+\Psi\left(E\right)-\Psi\left(S\right) \\ \nonumber&&
-\Delta^2V\left(S\right)\frac{\Psi\left(W\right)+\Psi\left(E\right)}{4}+{\cal O}\left(\Delta^4\right).
\end{eqnarray}
Here, we have the following points for the integration scheme: $N\equiv\left(u+\Delta,v+\Delta\right)$, $W\equiv\left(u+\Delta,v\right)$, $E\equiv\left(u,v+\Delta\right)$, and $S\equiv\left(u,v\right)$. This method was used a in a large number of works \cite{Konoplya:2014lha,Konoplya:2005et,Churilova:2021tgn,Aneesh:2018hlp,Qian:2022kaq,Varghese:2011ku} and proved its accuracy. Typical examples of time-domain profiles at $\ell =0$ and $\ell =1$ are shown in figs. 4 and 5. For $\ell =0$ the period of quasinormal ringing is much shorter than for higher multipoles, consisting usually only from a few oscillations. Therefore, it is difficult to extract the fundamental frequency with high precision. Nevertheless, comparison with accurate calculations by other methods for various black hole models says that even for  $\ell =0$ the relative error, when extracting  the fundamental mode from the time-domain profile, is  about $2-3\%$ which is more accurate than the 3d order WKB data. The numerical data for time-domain profiles are available from the author upon request.

As quasinormal modes have complex values, we will define the relative error between the accurate $\omega_{a}$ obtained by the time-domain integration (or higher order WKB data with Padé approximants) and approximate values $\omega$ given by analytical formulas as
\begin{equation}
E = |\frac{\omega_{a} - \omega}{\omega}|.
\end{equation}

In Tables I-VI above, it is evident that {\it for relatively small values of the dilaton parameter $b$}, the relative error of the analytic formula remains within $2-3\%$, even for the lowest multipole number $\ell = 0$ in the case of the scalar field, while for higher $\ell$ the error is smaller than one percent.At the same time, small values of the dilaton parameter is what expected from a quantum correction, because the perturbative correction should not change the geometry strongly. As $\ell$ increases, the analytic formula continues to exhibit reasonable accuracy, even for moderate dilaton values. A comparison of our data with previous significant results obtained in \cite{Fernando:2003wc,Shu:2004fj} using the third-order WKB approach reveals that the relative error produced by the third-order WKB formula reaches almost $7\%$ for $\ell=0$ scalar modes, even in the Schwarzschild case, as can be seen in table V. Consequently, for relatively small values of the dilaton parameter the analytic formula presented here proves to be more accurate than numerical data of \cite{Fernando:2003wc,Shu:2004fj} obtained by the 3d order WKB method.

\vspace{5mm}

\section{Conclusions}

An essential black hole solution arising in the low-energy limit of string theory is the dilaton black hole, initially found in \cite{Gibbons:1987ps,Garfinkle:1990qj}. In this paper, we derive the analytic expression for the quasinormal modes of the dilaton black hole, specifically for scalar and Dirac perturbations \cite{Gibbons:1987ps}.
Our results are compelling, exhibiting unexpectedly robust agreement when compared with the 6th order WKB formula employing Padé approximants, as well as with time-domain integration. Notably, the analytic formula surpasses the accuracy of certain previously published  numerical results \cite{Fernando:2003wc,Shu:2004fj} utilizing the lower order WKB formula for small and moderate values of the dilaton parameter. Our study opens avenues for further exploration, particularly in extending these calculations to analyze the grey-body factors of the dilaton black hole. Furthermore, we establish that the correspondence between eikonal quasinormal modes and null geodesics holds for test scalar and Dirac fields.

\vspace{4mm}

\begin{acknowledgments}
The author would like to acknowledge Roman Konoplya for valuable help with the time-domain integration technique.
\end{acknowledgments}

\bibliographystyle{unsrt}
\bibliography{bibliography}
\end{document}